\documentclass[usenatbib,conference]{basi}
\usepackage[T1]{fontenc}
\usepackage[british]{babel}
\usepackage[varg]{txfonts}
\usepackage{floatrow}
\newfloatcommand{capbtabbox}{table}[][\FBwidth]
\usepackage{rotating}
\usepackage{dcolumn}
\usepackage{ragged2e}
\usepackage{etoolbox}
\usepackage{array}
\newcolumntype{L}[1]{>{\raggedright\let\newline\\\arraybackslash\hspace{0pt}}m{#1}}
\newcolumntype{C}[1]{>{\centering\let\newline\\\arraybackslash\hspace{0pt}}m{#1}}
\newcolumntype{R}[1]{>{\raggedleft\let\newline\\\arraybackslash\hspace{0pt}}m{#1}}
\apptocmd{\frame}{}{\justifying}{} 

\usepackage{hyperref}
\usepackage{url}
\usepackage{graphicx}
\usepackage{natbib}
\usepackage{wrapfig,caption}
\usepackage{color,colortbl}

\bibpunct{(}{)}{;}{a}{}{,}
\usepackage{xcolor}
\definecolor{dark-red}{rgb}{0.5,0.15,0.15}
\definecolor{dark-blue}{rgb}{0.15,0.15,0.5}
\definecolor{medium-blue}{rgb}{0,0,0.5}
\definecolor{medium-red}{rgb}{1,0,0}
\hypersetup{
    colorlinks={true}, linkcolor={medium-red},
    citecolor={dark-blue}, urlcolor={medium-blue}
}

\graphicspath{{figures/}} 

\newcommand{\teff}{$T_{\scriptsize{\textrm{eff}}}$}
\newcommand{\logg}{\textrm{log g}}
\newcommand{\feh}{\textrm{[Fe/H]}}

\newcommand*\aap{A\&A}

\newcommand*\aj{AJ}

\newcommand*\apj{ApJ}

\newcommand*\apjs{ApJS}

\newcommand*\pasj{PASJ}

\begin{document}
\title[Estimating Stellar Atmospheric Parameters]{Estimating Stellar Atmospheric Parameters by Automated Methods Using SSLs}
\author[Sharma~et~al.]%
       {Kaushal~Sharma$^1$\thanks{email:\texttt{kaushalksharma1989@gmail.com}},
       H.~P.~Singh$^1$, A.~Kashyap$^1$ and P.~Prugniel$^2$\\
       $^1$ Department of Physics \& Astrophysics, University of Delhi, India\\
       $^2$ CRAL, Observatoire de Lyon, CNRS UMR 5574, 69561 Saint-Genis Laval, France}

\pubyear{2017}
\volume{14}
\pagerange{69--72}
\setcounter{page}{69}

\date{Received July 1st, 2017; accepted Oct 15th, 2017}

\maketitle
\label{firstpage}

\begin{abstract}
Libraries of stellar spectra, such as ELODIE \citep{PS01}, CFLIB \citep{valdes2004}, or MILES \citep{san2006}, are used for a variety of applications, and especially in modelling stellar populations (e. g. \citet{leborgne2004}). In that context, apart from the completeness and quality of these spectral databases \citep{singh2006}, the accurate calibration of stellar atmospheric parameters, temperature (\teff), surface gravity (\logg), and metallicity (\feh), is known to be critical \citep{prugniel2007,percival2009}. We discuss the technique of determining stellar atmospheric parameters accurately by `full spectrum fitting'.
\end{abstract}

\begin{keywords}
   atmospheric parameters, automated methods
\end{keywords}

\section{Introduction}\label{s:intro}

Significant progress in the fields of optical/spectroscopic techniques and observatory instrumentation has resulted in rapid increases in volumes of astronomical data. Various ground and space-based surveys\,-\,for instance, NASA's Hubble Space Telescope (HST), the Sloan Digital Sky Survey \citep[SDSS][]{york2000} and European Space Agency's Gaia mission \citep{perryman2001}\,-\,provide an enormous volume of data. Observatories equipped with next-generation spectrographs such as MOONS \citep{cirasuolo2011} have made data collection process more efficient. The growing size of observed data requires automated methods to analyse stellar spectra and provide their parameters. One such automated method is the `full spectrum fitting' technique, which uses a stellar spectral library (SSL) for estimating the atmospheric parameters (\teff, \logg{} and \feh).

Full spectrum fitting has been shown to be reliable in retrieving the atmospheric parameters for late A, F, G and early K spectral-type stars \citep{prugniel2011}, but for cooler stars (\teff{}$\,<\,4800$\,K) the reliability degrades. In this paper we present improved parameters of cool stars which were derived by full spectrum fitting. We also derive the parameters of 1065 cluster stars, belonging to 13 clusters, from SDSS. In Sec.~\ref{s:technique} we describe briefly the method of full spectrum fitting, along with modifications in the method that give better results for cool stars. The following section, Sec.~\ref{s:SDSS}, presents the determination of parameters for SDSS cluster stars, together with a comparative analysis of those parameters with ones in the literature.

\section{Full Spectrum Fitting Technique}\label{s:technique}

We used ULySS \citep{koleva2009} for employing the full spectrum fitting technique. It uses a $\chi^2$ minimisation algorithm for estimating parameters. A model spectrum is generated from a spectral library by using a `spectral interpolator'.

\begin{figure}
\includegraphics[scale=0.095]{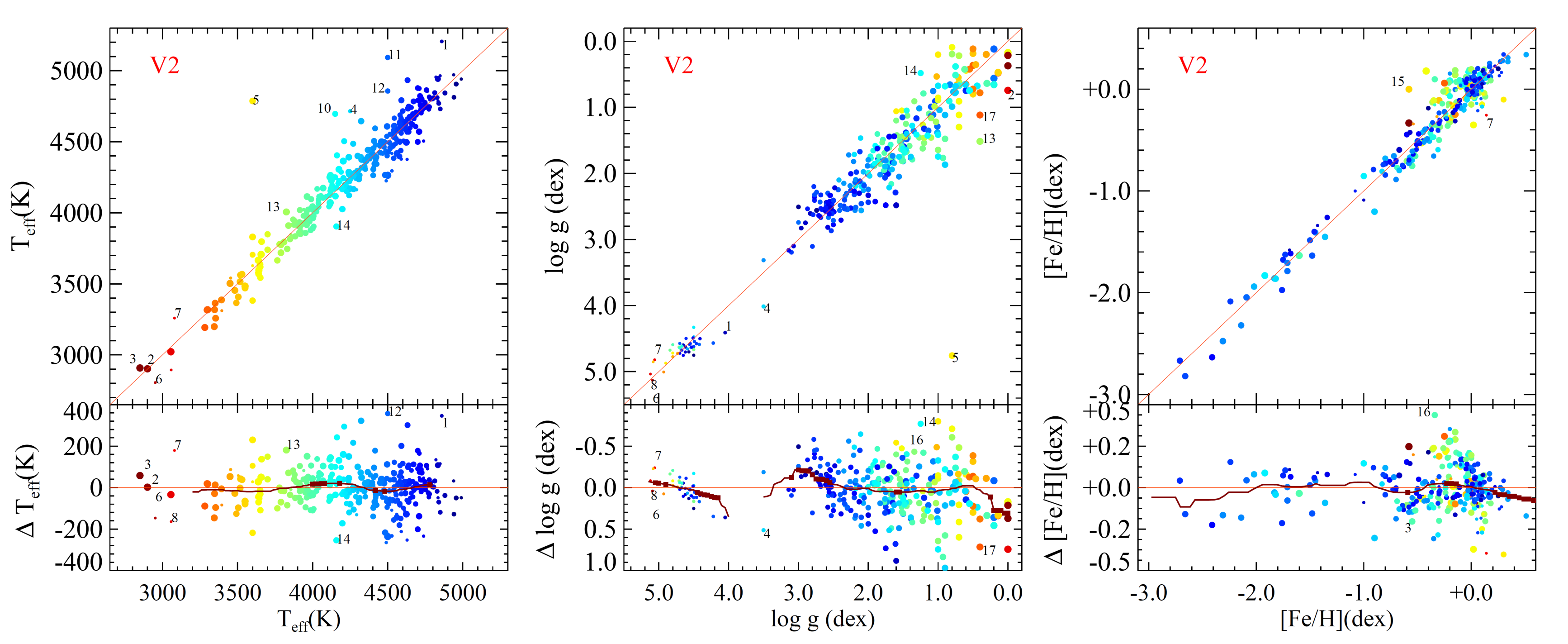}
\caption{Upper panels compare the derived parameters (ordinates) with the literature values (abscissae). The lower panels display the residuals.}\label{fig:comparison_v2}
\end{figure}

\textbf{Parameters of cool stars using revised MILES interpolator, V2}: 
To improve the estimation of parameters for cool stars, a sample of 331 stars (\teff{}$\,\leq\,$4800\,K) was chosen from MILES SSL. Significant biases in the re-derived parameters of those stars were detected with the MILES interpolator \citep[V1;][]{prugniel2011}. We improved that by correcting the input catalogue for systematics/biases, using a new polynomial (26 terms), and by supporting the extrapolation for cool stars by incorporating spectra of stars within the range M0$-$M4.5 from \citet{neves2013}. A new version of the interpolator \citep[V2;][]{sps2016} was derived. A comparison of our re-determined parameters with those in the literature is shown in Fig.~\ref{fig:comparison_v2}.

\section{Analysis of cluster stars from SDSS}\label{s:SDSS}

\begin{figure}
\includegraphics[scale=0.055]{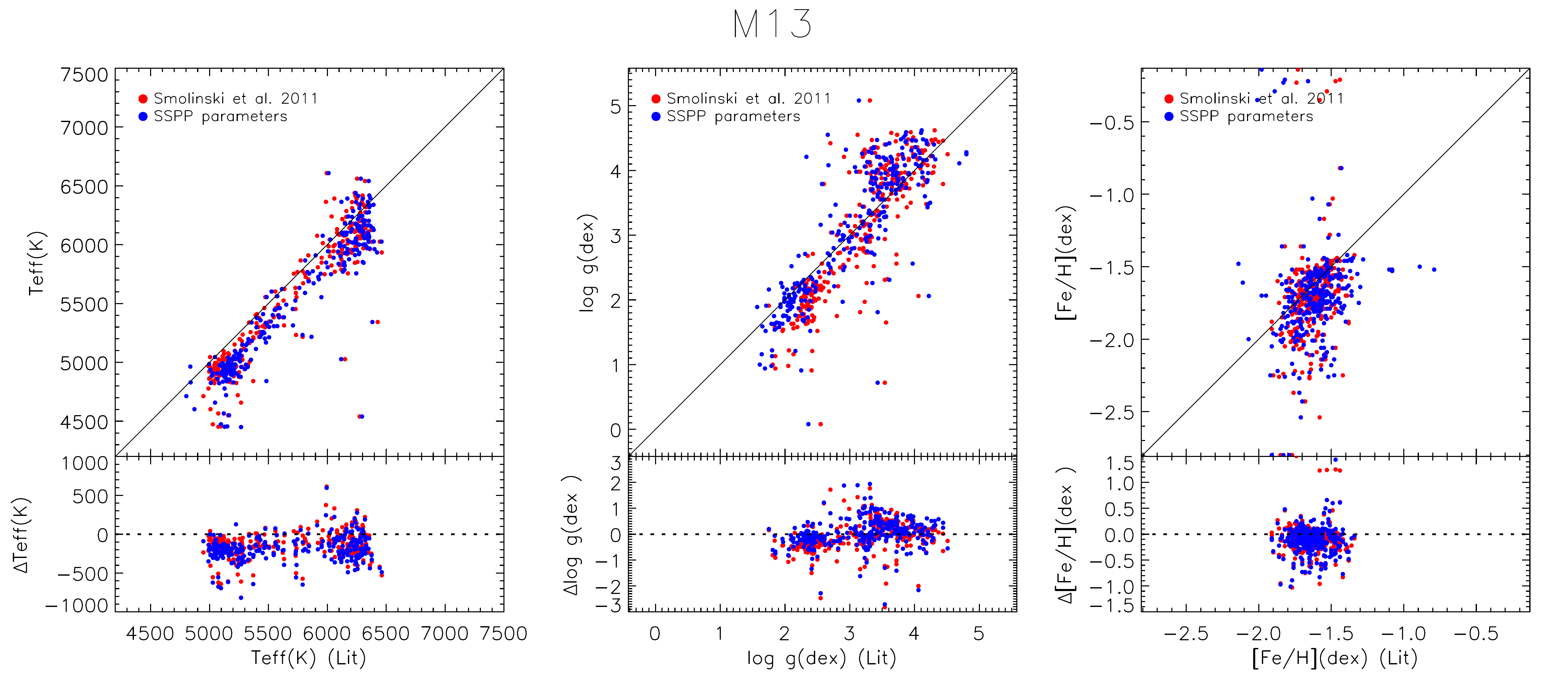}\\
\includegraphics[scale=0.055]{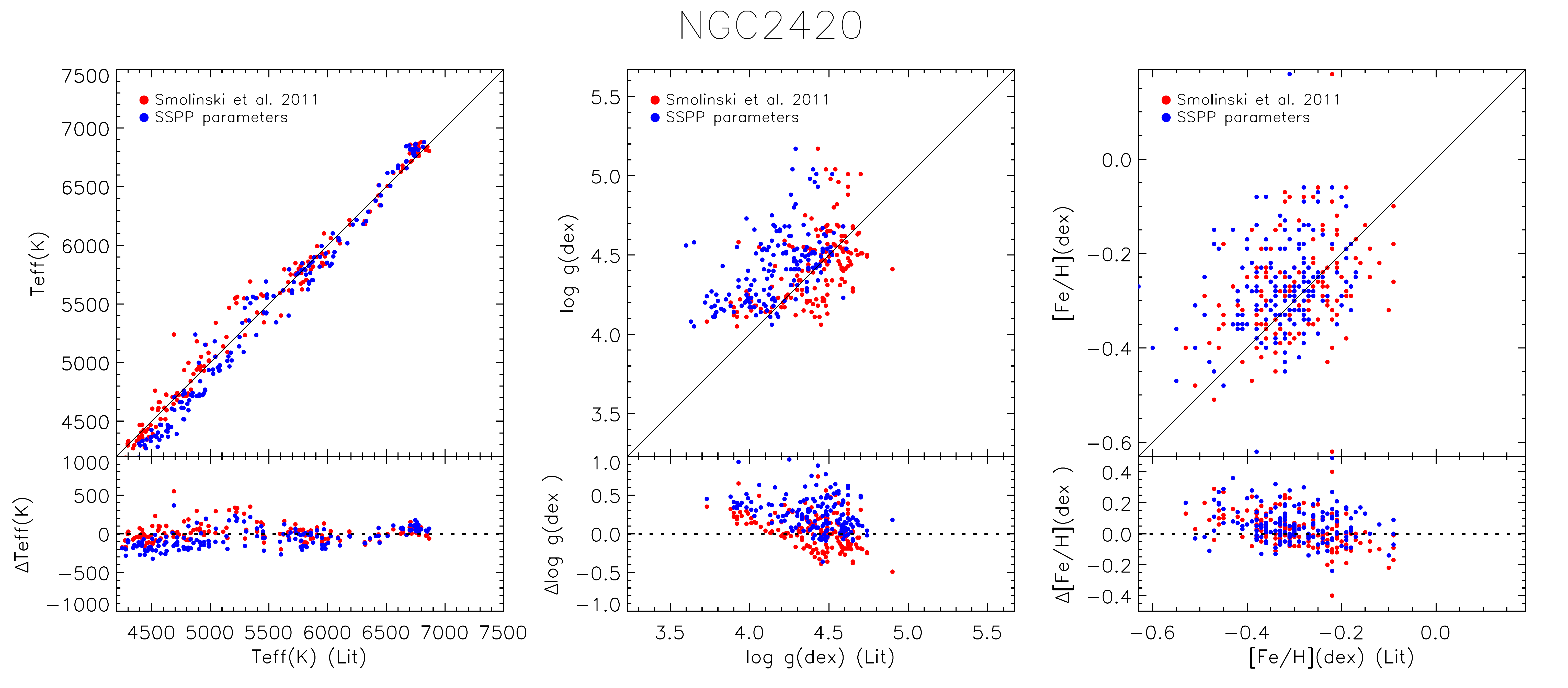}
\caption{Comparison of derived parameters with corresponding values from the literature for the two sample clusters.}\label{fig:cluster_comparison}
\end{figure}

1065 cluster stars belonging to 8 globular clusters (GCs) and 5 open clusters (OCs) were analysed to estimate their atmospheric parameters and cluster metallicities. The spectra were taken from SDSS DR12, and we used ULySS with the MILES V2 interpolator to determine the parameters. To cross-validate the method, we compared the derived parameters with those from \citet{smolenski2011} (Fig.~\ref{fig:cluster_comparison}). Statistics of the comparison are presented in Table~\ref{tab:sdss_cluster}. The results are also compared with the SDSS DR12 parameters that were obtained using the SEGUE Stellar Parameter Pipeline (SSPP) (Fig.~\ref{fig:cluster_comparison}). We plan to examine the systematic effects, and any dependence of
the parameters on S/N, etc., to improve the determinations further.

\begin{table}
\begin{center}
\scalebox{0.95}{
\begin{tabular}{llrrrrrrc}
\hline
Cluster & N & \multicolumn{2}{c}{$\Delta$\teff{}(K)} & \multicolumn{2}{c}{$\Delta$\logg{}(dex)} & \multicolumn{2}{c}{$\Delta$\feh{}(dex)} & Cluster \\ 
        &   & \multicolumn{1}{c}{$\mu$} &  \multicolumn{1}{c}{$\sigma$} & \multicolumn{1}{c}{$\mu$} &  \multicolumn{1}{c}{$\sigma$} & \multicolumn{1}{c}{$\mu$}  &  \multicolumn{1}{c}{$\sigma$} & Type \\
\hline
M2        & 70   &  -221 &    308 &  -0.22 &   0.70 &  -0.13 &   0.37 & GC\\
M3        & 77   &  -155 &    278 &  -0.41 &   0.61 &  -0.02 &   0.32 & GC\\
M13       & 293  &  -146 &    339 &  -0.07 &   0.59 &  -0.10 &   0.33 & GC\\
M15       & 98   &  -271 &    335 &   0.36 &   1.00 &  -0.04 &   0.35 & GC\\
M35       & 29   &  -128 &    264 &  -0.05 &   0.27 &   0.07 &   0.09 & OC\\
M53       & 17   &  -107 &    981 &   0.13 &   0.60 &   0.16 &   0.73 & GC\\
M67       & 74   &   -77 &    144 &  -0.16 &   0.27 &  -0.04 &   0.08 & OC\\
M71       & 17   &  -200 &     96 &  -0.61 &   0.20 &  -0.11 &   0.11 & GC\\
M92       & 58   &  -342 &    537 &  -0.25 &   0.99 &  -0.22 &   0.40 & GC\\
NGC2158   & 62   &  -330 &    209 &   0.01 &   0.40 &  -0.08 &   0.07 & OC\\
NGC2420   & 164  &     7 &    111 &   0.01 &   0.21 &  -0.04 &   0.10 & OC\\
NGC5053   & 16   &  -334 &    214 &  -0.49 &   0.37 &  -0.07 &   0.49 & GC\\
NGC6791   & 90   &   -66 &    149 &  -0.38 &   0.31 &  -0.07 &   0.05 & OC\\

\hline
\end{tabular}
}
\caption{Statistics from comparing the derived parameters with those of \citet{smolenski2011}.}\label{tab:sdss_cluster}
\end{center}
\end{table}

\section*{Acknowledgements}

HPS is grateful to organisers for generous support for attending the meeting.


\label{lastpage}
\end{document}